\documentclass[twocolumn,showpacs,preprintnumbers,amsmath,amssymb]{revtex4-1}

\usepackage{graphicx}
\usepackage{dcolumn}
\usepackage{bm}
\usepackage{amsmath}
\usepackage{color}
\usepackage{soul}

\begin{document}

\title{Directional lasing at high oblique angle from metasurface exhibiting multivalley dispersion}

\author{H. S. Nguyen$^1$}
\author{F. Dubois$^1$}
\author{T. Deschamps$^1$}
\author{S. Cueff$^1$}
\author{P. Regreny$^1$}
\author{J-L. Leclercq$^1$}
\author{C. Seassal$^1$}
\author{X. Letartre$^1$}
\author{P. Viktorovitch$^1$}

\email{hai-son.nguyen@ec-lyon.fr}
\affiliation{$^1$Institut des Nanotechnologies de Lyon, INL/CNRS, Universit\'e de Lyon, 36 avenue Guy de Collongue, 69130 Ecully , France}

\date{\today}

\begin{abstract}
We report on a metasurface laser emitting at the valley extremum of a multivalley energy-momentum dispersion. Such peculiar dispersion shape is obtained by hybridizing high quality factor photonic modes of different symmetry and opposite effective mass. The lasing effect takes place at high oblique angle ($\sim$20 degrees), in the telecom wavelength range ($\sim$1580 nm), on silicon substrate and operating at room temperature. Our results show the potential of multivalley dispersion for micro-laser in integrated photonic and  beam-steering applications. It also opens the way to study various features of valleytronic physics such as spontaneous momentum symmetry breaking, two-mode squeezing and  Josephson oscillation in momentum space.
\end{abstract}

\maketitle
Manipulating emission is one of the most fundamental features of photonic research. In the wake of Purcell's seminar paper~\cite{Purcell1946}, the usual way to control spontaneous emission is to modify the Local Density of Optical States (LDOS) of electromagnetic environment where emitters are embedded. This has been widely achieved in the past by implementing emitters in microcavity~\cite{Gerard1998}, photonic crystal~\cite{Lodahl2004,Englund2005,Fujita2005,Noda2007} and plasmonic structures~\cite{Anger2006,Kuhn2006,Tanaka2010,Akselrod2014}. The Purcell theory considers mostly the near-field effect of light-matter interaction for accelerating/inhibiting the spontaneous emission, and the early works do not pay much intention to the far-field pattern of light emission~\cite{Pelton2015}. 

Otherwise, modern applications such as LIDAR (light detection and ranging), micro-display and LIFI (optical wireless communication) demand not only efficient light emission (i.e. high rate, low non-radiative losses and high extraction efficiency) but also the possibility to engineer and control its direction. Over the last few years, resonant metasurface~\cite{Kuznetsov2016,Staude2017} has emerged as a very promising platform to develop directional emitting devices. Indeed, these metamaterials, composed of individual or collective subwavelength resonators, make it possible to combine the Purcell effect in the near-field scale, for boosting light emission, with the control of energy-momentum dispersion in the far-field scale, for dictating the emitting direction. To obtain this double effects, the most common method is to embed emitters in the vicinity of a single or few coupled metallic resonators for enhancing the light emission, then tuning the geometry of resonators to tailor the far-field pattern~\cite{Coenen2011,Aouani2011,Aouani2011a,Yousefi2012,Coenen2014,Verre2016,Lindfors2016,Pantoja2017,Sattari2017}.  However, these designs exhibit high losses due to metallic structures, and the research trend is going towards all-dielectric metasurfaces~\cite{Kuznetsov2016,Staude2017}.  Indeed, dielectric structures provide not only high quality factor resonances, but also benefit dispersion engineering concepts emerged in the domain of photonic crystal. In this approach, several groups have reported on directional emission by coupling emitters to periodical lattice of particular dispersion: fast-modes~\cite{Wu2014,Hoang2017}, slow-light modes~\cite{Chen2017,Zhang2017,Vaskin2018} and Dirac cones~\cite{Moitra2013}.  Very recently, Ha \textit{et al} has demonstrated the first off-axis vertical lasing by introducing a leaky channel to the dispersion of a bound state in the continuum~\cite{Ha2018}. Although very promising, this mechanism requires a trade-off between the oblique angle emission and quality factor, thus demand high threshold for lasing at high oblique emission.

In this letter, we demonstrate experimentally a directional lasing effect based on dielectric metasurface exhibiting multivalley dispersion. Such peculiar energy-momentum dispersion is obtained by hybridizing two high quality factor resonant modes of different symmetry and of opposite photonic mass. Angle-resolved photoluminescence experiment is performed to study the far-field pattern of the metasurface, using InAsP quantum wells as gain material. At low power excitation, W-shaped multivalley is revealed in the farfield photoluminescence spectrum. Such dispersion is perfectly reproduced by numerical simulation and analytical calculation. When increasing the pumping power, lasing effect is observed in the valley minima, corresponding to 20 degrees off-axis angle. Such micro-laser emits light in the telecom range, on silicon substrate and operating at room temperature.  The design can be easily adapted to different family of semiconductor and emitters for applications in integrated photonic and to study the valleytronic physics.
\begin{figure}[htb]
\begin{center}
\includegraphics[width=8.5cm]{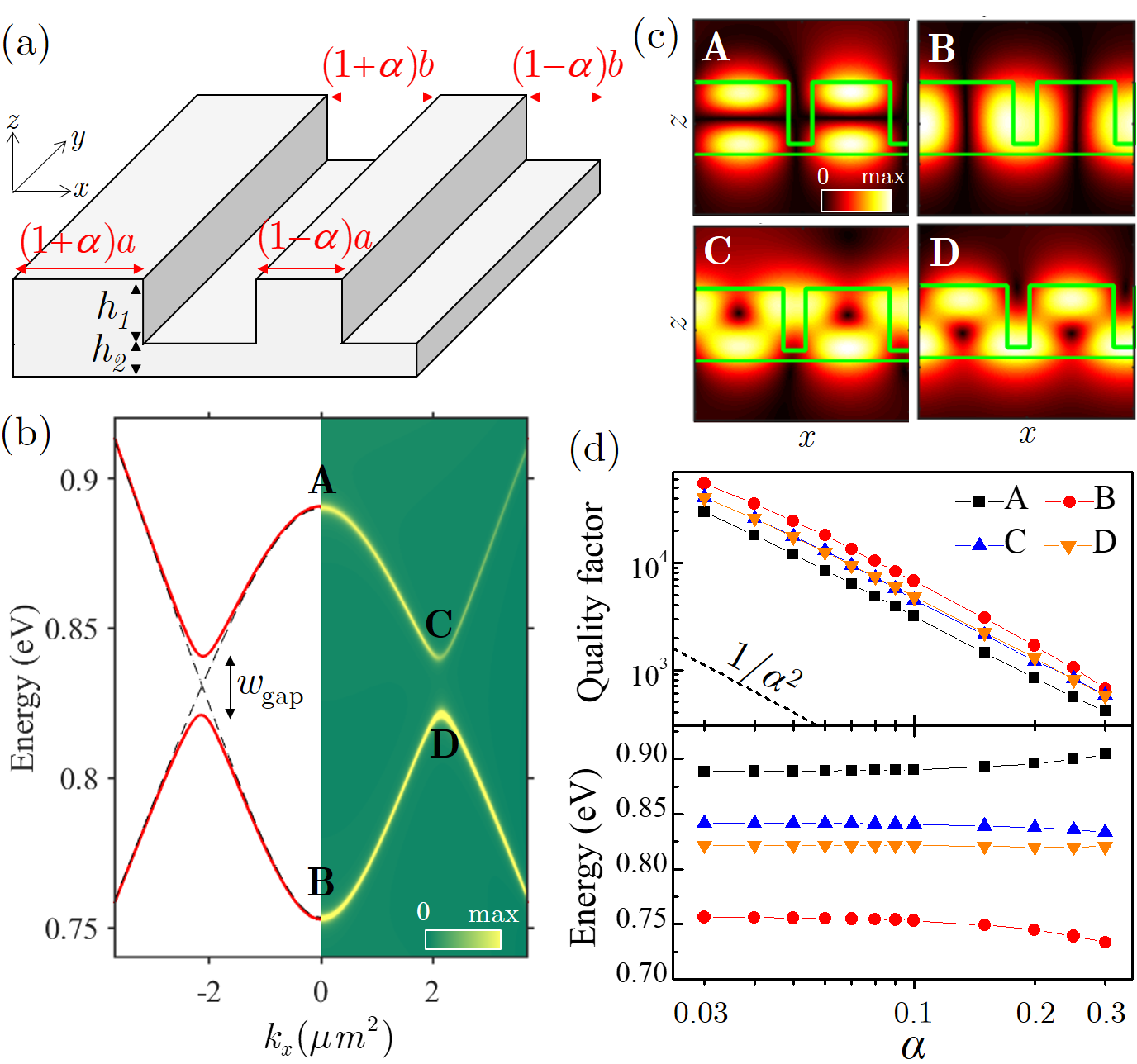}
\end{center}
\caption{\label{fig1}{(a) Sketch of the unit cell of the metasurface providing multivalley dispersion of high quality factor photonic modes. (b) \textit{Right panel:} Numerical simulation of the energy-momentum dispersion for the metasurface made from a dielectric of refractive index $n=3.54$, and as parameters: $a=256\,nm$, $b=64\,nm$, $h_1=300\,nm$, $h_2=50\,nm$, $\alpha=0.1$. \textit{Left panel:} calculation of the dispersion without the coupling (black dashed line) and with the coupling (red line), using the analytical model given by eq.~2. (c) Distribution of the electric field at different extrema of the dispersion diagram. (d) Numerical simulations of the quality factor (upper panel) and the energy (lower panel) at different extrema of the dispersion as a function of the double-period peturbation $\alpha$.}}
\end{figure}

The design of our resonant metasurface consists of a dimerized one-dimensional lattice made from high-contrast index material. The unit cell design is presented in Fig~1(a). It has been recently demonstrated that such architecture exhibits many degree of freedom to engineer the energy-momentum dispersion and the quality factor of photonic modes~\cite{Nguyen2018}. In the following we will discuss how such design can provide multivalley dispersion of photonic modes with high quality factor:

\underline{\textit{Multivalley dispersion:}} The vertical symmetry breaking due to the presence of a ``residual'' slab [i.e. the unpatterned layer of thickness $h_2$ in Fig.~1(a)] makes it possible to couple modes of opposite vertical symmetry, thus changes drastically the photonic dispersion from the one of symmetric design~\cite{Nguyen2018}. To obtain multivalley dispersion, we harness the hybridization between two TE-polarized modes $E_y$ of conventional parabolic dispersions with opposite curvature (i.e. photonic mass) [see Fig~1(b)]. Such coupling results in an anticrossing effect, corresponding to valley formation [point C and D in Fig~1(b)] in a W-shaped and M-shaped dispersion. The distribution of the electric field at different extrema are shown in Fig~1(c), confirming that the mixed states (C and D) are indeed the hybridization of an odd mode (A) and an even mode (B). In an analytical model of mode coupling, the Hamiltonian of the system is given by~\cite{Zhen2015}: 
\begin{equation}
    H(k_x)=
    \begin{bmatrix}
    E_{o} - i\gamma_{o} & V-i\sqrt{\gamma_o\gamma_e} \\
    V-i\sqrt{\gamma_o\gamma_e} & E_{e} - i\gamma_e \\
\end{bmatrix}
\end{equation}
where $E_{o(e)}$, depending on $k_x$, correspond to the parabolic dispersions of the uncoupled odd(even) modes; $\gamma_{o(e)}$ are their radiative losses; and $V$ is the coupling strength induced by the vertical symmetry breaking. By diagonalizing the previous Hamiltonian, we obtained two hybdrid-states:
\begin{multline}
    E_\pm - i\gamma_\pm = \frac{E_{o}+E_{e}}{2} - i\frac{\gamma_{o}+\gamma_{e}}{2} \\  \pm\frac{\sqrt{\big[(E_{o}-E_{e})-i(\gamma_{o}-\gamma_{e})\big]^2 + 4(V^2-\gamma_o\gamma_e)}}{2}
\end{multline}
Several remarks can be deduced from this very simple derivation: \textbf{i) Far from the anticrossing point} (i.e. $\left|E_o-E_e\right|\gg V$), the coupling effect is negligible. The energy and losses are then given by the one of uncoupled modes. \textbf{ii) At the anticrossing point} (i.e. $E_o=E_e$), using the real part of Eq.~2, the gap-opening would amount to  $w_{gap}=\sqrt{4V^2-(\gamma_o+\gamma_e)^2}$. This gap requires a coupling strength being  more efficient than the radiative losses: $V>(\gamma_o+\gamma_e)/2$.  On the other hand, using the imaginary part of Eq.~2, we note that the losses would be perfectly balanced between the two hybrid modes: $\gamma_+=\gamma_-=(\gamma_o+\gamma_e)/2$. In term of quality factor, this balance corresponds to $\frac{1}{Q_+}=\frac{1}{Q_-}=\frac{1}{2}(\frac{1}{Q_o}+\frac{1}{Q_e})$ 

\underline{\textit{Control of high quality factor:}} To obtain modes of high quality factor, the dimer design with double-period perturbation $\alpha\ll1$ is adopted~\cite{Nguyen2018}. Indeed, the two modes discussed above were originally at the edge of the Brillouin zone ($k_x=\frac{\pi}{a+b}$) and below the lightline of the monomer design (i.e. $\alpha=0$). They were thus lossless and could not radiate to the free space. By implementing the double-period perturbation, we obtain an unit cell of period twice larger, thus a Brillouin zone twice smaller than the one of the monomer. The previous modes are then folded to $k_x=0$ and start radiating to the free space. As consequence, the quality factor of the hybrid modes can be finely tuned by $\alpha$, following a very simple law: $Q\propto 1/\alpha^2$~\cite{Nguyen2018}. To illustrate this effect, the quality factor at the extrema of Fig.~1(b) are extracted via numerical simulations and reported in the upper panel on Fig.~1(d). The simulations evidence the dependence in $1/\alpha^2$ of the quality factors. Moreover, the balance of radiative losses at the valley extrema are also verified, with $\frac{1}{Q_C}=\frac{1}{Q_D}=\frac{1}{2}(\frac{1}{Q_A}+\frac{1}{Q_B})$. Furthermore, as shown in the lower panel of Fig.~1(d), when modifying $\alpha$, the spectral position of extrema stay at the same values (for $\alpha<0.1$). So the double period perturbation allows for a fine tuning of the quality factor while unchanging the farfield pattern of the metasurface. Finally, we note that $Q_C \approx 1.5Q_A$ for every values of $\alpha$, thus if a gain medium is inserted in the spectral range of the W-shaped dispersion, the lasing effect will be favoured to take place at the valley extremum (point C) instead at $k_x=0$ (point A). 

Based-on the dimer design discussed above, the structure of the fabricated sample is presented in Fig.~2(a). Scanning Electron Microscopy (SEM) images of the metasurface ($30\times30\,\mu m^2$) are shown in Fig.~2(b). The unpatterned slab (i.e. ``residual'' layer to break the vertical symmetry of the design) consists of an InP layer which was growth by Molecular Beam Epitaxy (MBE) on sacrificial buffer, then transfered by molecular bonding to $2\,\mu m$ of $SiO_2$ on Silicon substrate. As active materials, four InAsP quantum wells are embedded within the InP layer during the MBE growth. The patterned layer of dimerized lattice is made from amorphous silicon. This layer was first deposited by Plasma-Enhanced Chemical Vapor Deposition (PECVD) deposition. Then the patterning was performed via electronic beam lithography and plasma-assisted dry etching. In the same notations as the design in Fig~1(a), the geometrical parameters of the fabricated sample are: $h_1=250\,nm$, $h_2=245\,nm$, $a=189\,nm$, $b=171\,nm$ and $\alpha=0.1$. 
\begin{figure}[htb]
\begin{center}
\includegraphics[width=8.5cm]{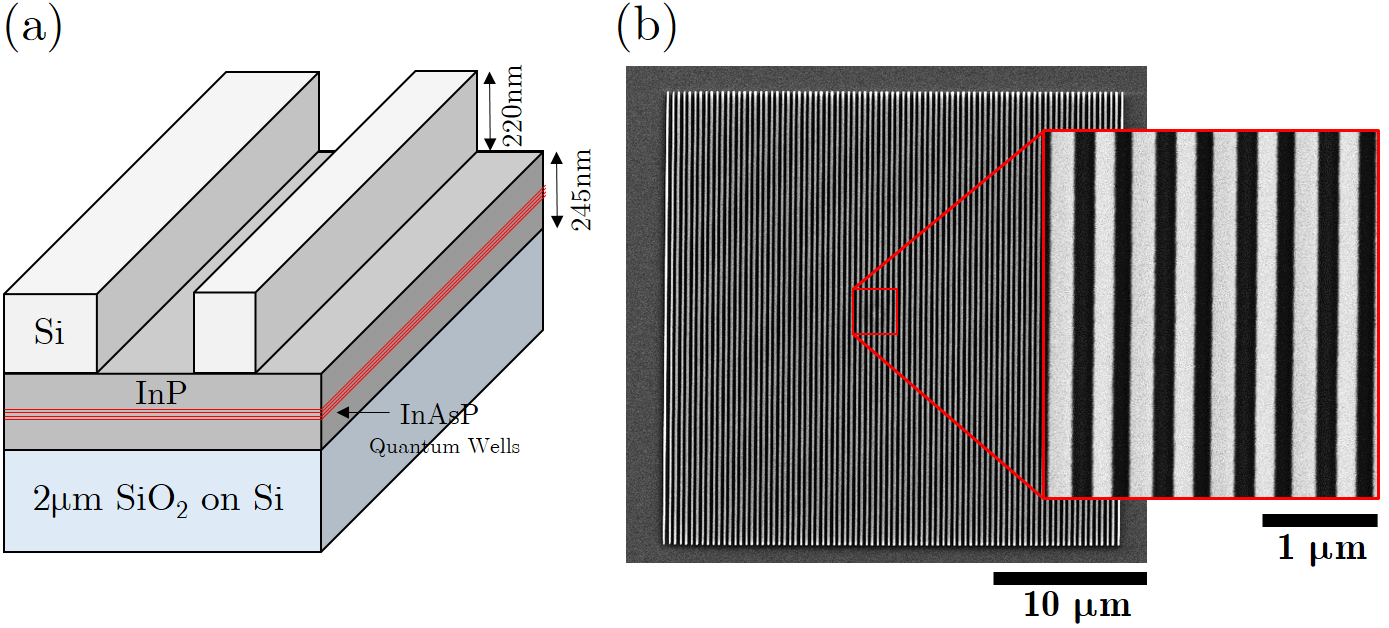}
\end{center}
\caption{\label{fig2}{(a) Sketch of the unit cell of the fabricated sample. (b) SEM images of the metasurace structure on the fabricated sample.}}
\end{figure}
\begin{figure}[htb]
\begin{center}
\includegraphics[width=8.5cm]{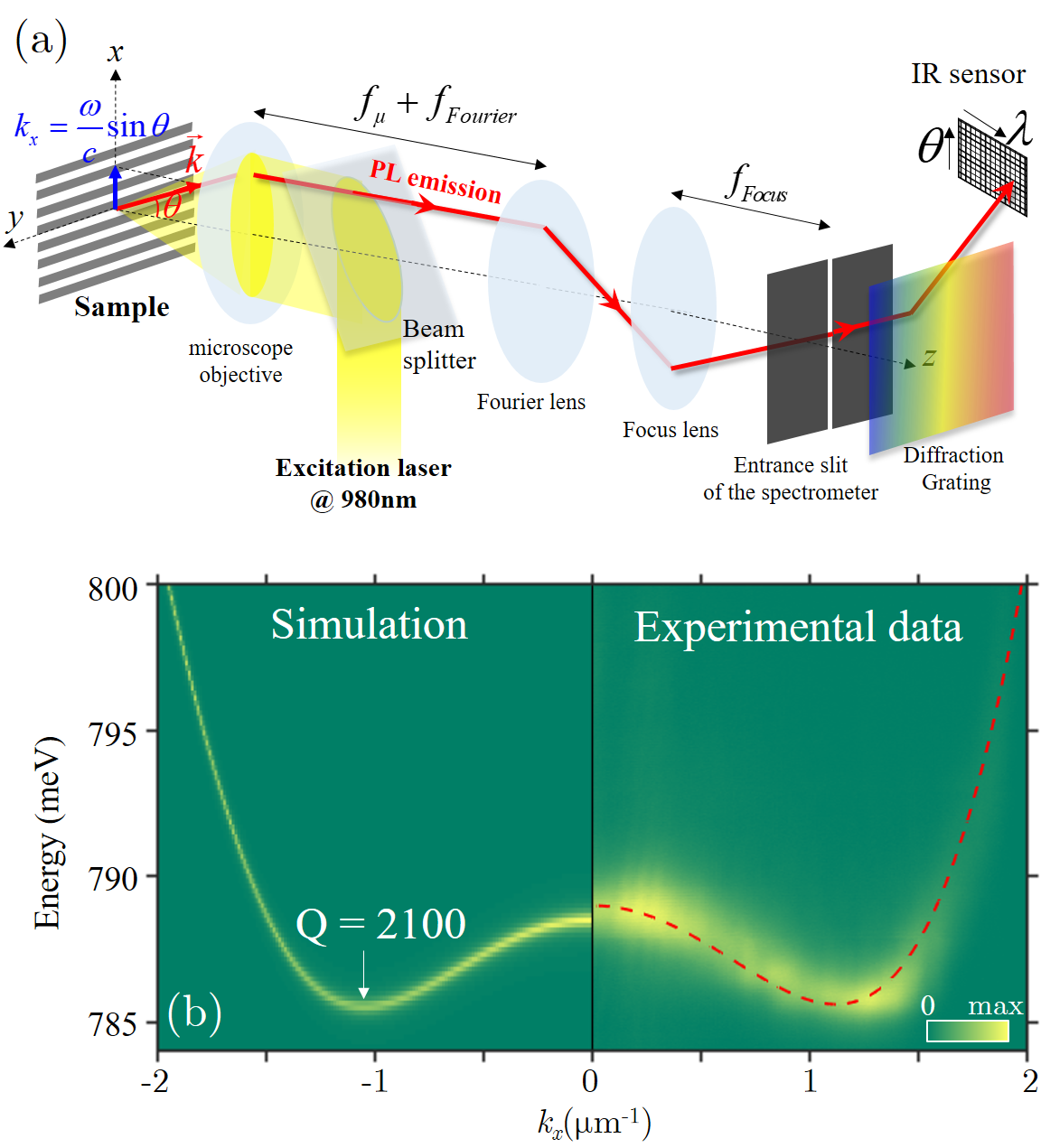}
\end{center}
\caption{\label{fig3}{(a) Sketch of the experimental setup for angle-resolved photoluminescence measurements. (b) Experimental results (right panel) and numerical simulation (left panel) of the photonic dispersion obtained from angle-resolved photoluminescence. The red-dashed line corresponds to the analytical fitting of the photonic dispersion.}}
\end{figure}

Angle-resolved photoluminescence experiment is performed to obtain the energy-momentum dispersion of the fabricated metasurface. The experimental setup is shown in Fig.~3(a). The excitation source is a 980$\,$nm laser which can be used in two modes: continous-wave (cw) or picosecond pulsed (pulse width of 50ps, repetition rate of 80 Mhz). The excitation laser is focused onto the metasurface via a microscope objective (x50, NA=0.42), forming a spot-size of $2-3\,\mu m$. Photoluminescence of InAsP quantum wells are collected via the same objective and projected onto the entrance slit of a spectrometer. The sample orientation is aligned so that the corrugated direction (i.e. x-axis) is along the entrance slit. A set of lenses are carefully used so that the projection of the photoluminescence image is in the Fourier space (i.e. $k_x$ and $k_y$). The output of the spectrometer is coupled to a CMOS sensor of InAs array (640x512). Thus the image obtained from the camera give us directly the energy-momentum dispersion diagram along $k_x$. A multivalley dispersion of W-shaped is clearly evidenced when pumping the metasurface by cw excitation [see fig.~3(a)]. The experimental dispersion is perfectly reproduced by numerical simulations and also well fitted by the analytical coupled-mode theory presented previously. We note that the measured mode is much broader than the theoretical prediction ($Q_{theory}=2100$ at the valley minimum) due to the absorption of the quantum wells. 
\begin{figure*}[htb]
\begin{center}
\includegraphics[width=16cm]{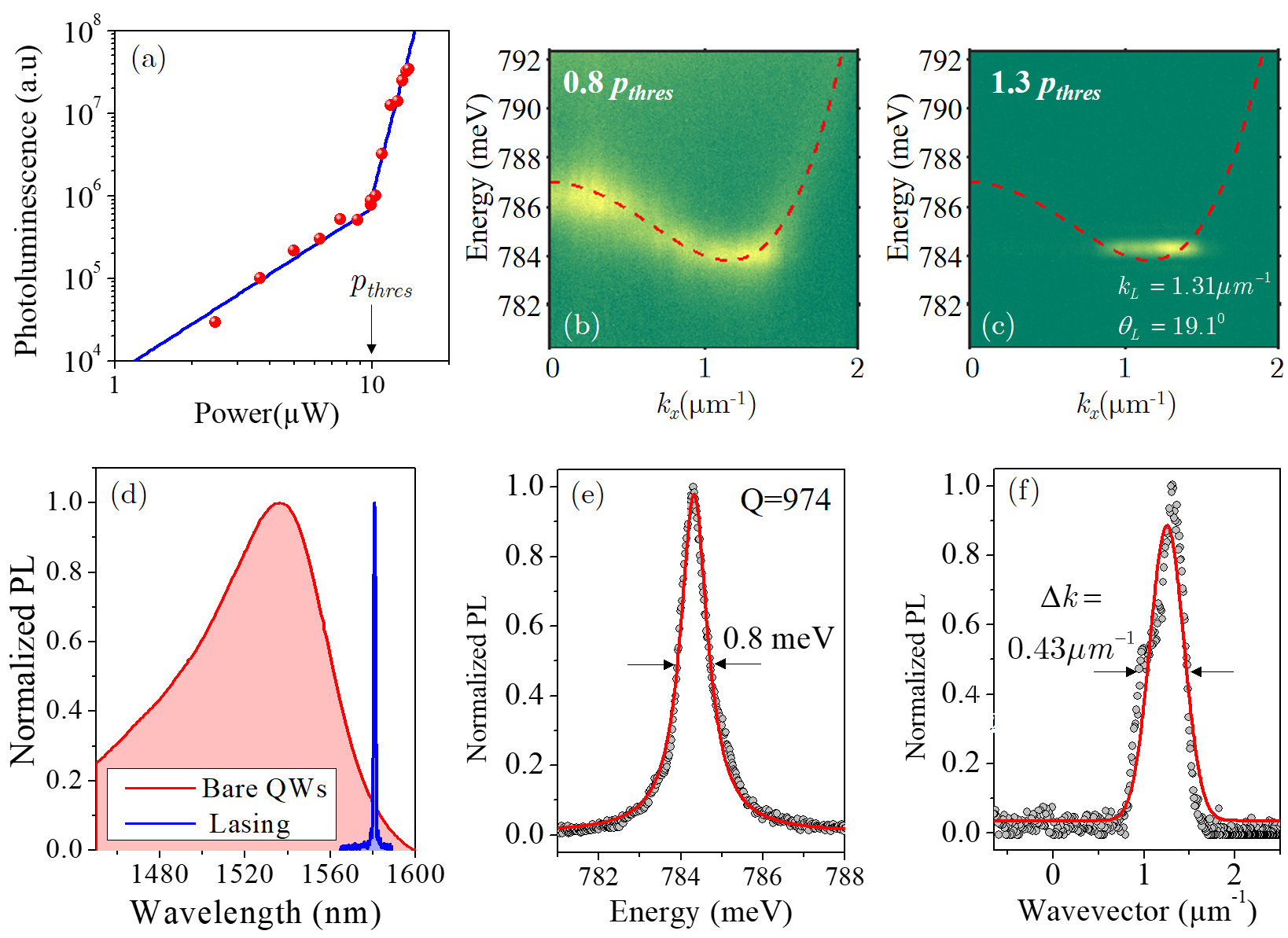}
\end{center}
\caption{\label{fig4}{(a) Integrated photoluminescence intensity as a function of the pump power under pulsed excitation. (b,c) Angle-resolved photoluminescence measurements when pumping (a) below, (b) above threshold. The red-dashed line corresponds to the analytical fitting of the photonic dispersion. (d) Emission spectrum of the spontaneous emission of InAsP quantum wells (red line) and of the lasing emission at pumping power $1.3p_{thres}$ (blue line). (e) Emission spectrum at pumping power $0.95p_{thres}$ and at emission angle of 19.1°. The red line is a Lorentzian fit of $0.8\,meV$ linewidth, corresponding to a quality factor of $Q=974$. (f) Intensity of lasing emission at pumping power $1.3p_{thres}$ as a function of wavevector. The red line is a Gaussian fit of $0.43\mu m^{-1}$ linewidth.}}
\end{figure*}

To study the lasing effect, photoluminescence experiment at different pumping power of pulsed excitation has been performed. Figure.~4(a) reports the integrated intensity as a function of excitation power, showing clearly a lasing effect with pumping threshold of $p_{thres}\approx\,10\,\mu W$, measured at the output of microscope objective. This average power corresponds to a pumping fluence of $2.5\,\mu J/cm^2$.  Angle-resolved photoluminescence is monitored when increase the pumping power from below threshold [i.e. $0.8\,p_{thres}$, see Fig.~4(b)] to above threshold [i.e. $1.3\,p_{thres}$, see Fig.~4(c)]. Below threshold, similar to the case of cw pumping, the spontanous emission of InAsP quantum wells ``populates'' the whole dispersion branch, leading to a very broadband emission of more than 200\,nm. Most interestingly, above threshold, we clearly observe the stimulation of emission to the valley point of W-shaped dispersion ($k_L=1.31\,\mu m^{-1}$). This lasing effect is associated to an emission at 1581\,nm (i.e. 784\,meV), emitting along a very high oblique angle $\theta_L=19.1°$, and within a sharp linewidth $0.8\,nm$ (i.e. $0.4\,meV$ at $1.3\,p_{thresh}$) when compared to the one of spontanous emission [see Fig.~4(d)]. Although the quality factor estimated from the lasing linewidth is as high as $Q_{lasing}=1975$ and very close to the one theoretically predicted ($Q_{theory}=2100$), this value is in fact limited by dephasing processus in the system, thus not representative to the real quality factor of our photonic mode. The correct method to estimate the quality factor is to extract the linewidth when the pumping power is slightly bellow threshold. Indeed, this pumping condition corresponds approximatively to the transparency regime of the gain material, in which there is no more broadening due to quantum well absorption. Figure.~4(e) depicts the emission spectrum at $\theta_L$ when pumping at $0.95\,p_{thres}$. From the extracted linewidth ($0.8\,meV$), we estimate a quality factor of $Q=974$. The fact that this experimental value is inferior than the one of the theoretical prediction is likely due to the fabrication imperfection. Finally, to characterize the directivity of our lasing emission, the emission intensity as a function of the in-plan wavevector, when pumping above threshold, is reported in Fig.~4(f). By fitting the experimental data with a Gaussian distribution, we extract an emission directivity within $\Delta k=0.43\,\mu m^{-1}$. This value is in good agreement ($\Delta_k.\Delta_x\approx 1$) with the size of the cavity effect induced by the pump of a $2-3\,\mu m$ spot-size. 

We highlight that the conception of our metasurface makes it possible to tailor separately the dispersion shape and the quality factor, thus suggests a unique scheme to obtain simultaneously valley extrema at high momentum and high quality factor. Such scenario leads to the possibility of directional lasing at arbitrary angle with very low threshold power. As matter of fact, for the same angle range, the threshold of our directional laser is one order of magnitude lower than one reported in the literature~\cite{Ha2018} which requires a trade-off between emission angle and quality factor.

Furthermore, it is possible to extend the photonic concept of our metasurface to polaritonic physics when light-matter interaction takes place in the strong coupling regime~\cite{Weisbuch1992}. This will be achieved by either working at cryogenic temperature or with quantum wells of high-binding energy excitons (GaN, ZnO, transition metal dichancogenides and perovskites)~\cite{Sanvitto2016}. As a matter of fact, the multivalley dispersion and valley lasing demonstrated in this work are the most important building blocks to study polariton valleytronic physics~\cite{Sun2017,Karpov2018}. Indeed, it has been recently predicted that strong coupling regime with W-shaped dispersion would lead to Bose Einstein condensation at the valley extrema~\cite{Karpov2018}. The formation of such macroscopic bosonic states at the valley extrema will pave the way to study spontaneous symmetry breaking, two-mode squeezing~\cite{Sun2017} and Josephson oscillation in momentum space~\cite{Zheng2018}.

In conclusion, by engineering multivalley dispersion of high quality factor, we have demonstrated experimentally a metasurface laser which emits coherent light of telecom wavelength along high oblique angle, on silicon substrate and operating at room temperature. The design of our devices provides a high degree of freedom for choosing the lasing direction and the quality factor of the photonic mode, thus very promising for integrated beam steering system of LIDAR and LIFI applications. Moreover, our design can be easily extended to other material family, and perfectly adapted for operation in the strong coupling regime of light-matter interaction for studying the valleytronics physics of  Bose Einstein condensation. 

The authors would like to thank the staff from the NanoLyon Technical Platform for helping and supporting in all nanofabrication processes. This work is partly supported by the French National Research Agency (ANR) under the projects PICSEL, SNAPSHOT and POPEYE.

\bibliography{Ref_Wlasing}

\end{document}